\documentclass[showpacs,preprint]{revtex4}

\usepackage{graphicx}
\usepackage{subfigure}
\usepackage{amssymb}
\usepackage{amsmath}
\usepackage{bm}
\usepackage{latexsym}
\usepackage{hyperref}

\newcommand{\Tr}{{\text Tr}}
\begin{document}

\title{Polynomial scheme for time evolution of open and closed
quantum systems}
\author{Jun Jing\footnote{Email: jingjun@sjtu.edu.cn}, H. R. Ma }
\affiliation{Institute of Theoretical Physics,
Shanghai Jiao Tong University\\
800 DongChuan Road, MinHang, Shanghai 200240, China}
\date{\today}

\begin{abstract}
Based on the generation function of Laguerre polynomials, We
proposed a new Laguerre polynomial expansion scheme in the
calculation of evolution of time dependent Schr\"odinger equation.
Theoretical analysis and numerical test show that the method is
equally as good as Chebyshev polynomial expansion method in
efficiency and accuracy, with extra merits that no scaling to
Hamiltonian is needed and wider suitability.
\end{abstract}

\maketitle

\section{Introduction}
The studies of open quantum systems have a long history
\cite{weiss}. There has been renewed interest in recent years due to
the developments of the conception and possible realization of
quantum communication and quantum computation \cite{breuer,nielson}.
A key concept in the open quantum system study is the decoherence of
a quantum system interacting with environments, which plays a very
important rule in almost all phenomena in the quantum devices used
in quantum computation and quantum communication
\cite{Joos,Zurek1981,Zurek2003}. It has been shown that the states
of an open quantum system will finally relax into a set of ``pointer
states'' in the Hilbert space \cite{Zurek1981} by decoherence, i.e.
for a quantum system prepared in a linear superposition of its
eigenstates, interaction of the system with its environment results
in a decay from the system's initial pure state,
$\rho_s(t=0)=|\psi_0\rangle\langle\psi_0|$, to a mixed state,
$\rho_s(t>0)=\sum_ip_i\rho_i$, $\sum_ip_i=1$. To be specific, an
arbitrary initial state of the system plus the environment may be
written as:
\begin{equation}\label{equ1}
|\psi(t=0)\rangle=(\sum_{n}C_{n}|n\rangle)\otimes|\psi_e\rangle,
\end{equation}
where the set $|n\rangle$ stands for the eigenstates of the system
and $|\psi_e\rangle$ is the initial state of the environment. This
state at time $t$ larger than the decoherence time $\tau_d$ evolved
to a mixed state, which may be expanded as:
\begin{equation}\label{equ2}
|\psi(t)\rangle=\sum_{m}C_{m}(t)(|m\rangle\otimes|e_{m}\rangle).
\end{equation}
Here, the set of states $|m\rangle$ are the so-called pointer states
of the system \cite{Dobrovitski1,Gardiner,Gottfried}, and
$|e_{m}\rangle$ are the corresponding states of the environment that
entangled with $|m\rangle$\cite{Braginsky}. A convenient way to
represent the system interacting with the environment is the reduced
density matrix, defined as
\[
\rho_{s}=\Tr_{e}\left(|\psi(t)\rangle\langle\psi(t)|\right),
\]
where $\Tr_{e}$ means tracing over the environment degrees of
freedom. The evolution from (\ref{equ1}) to (\ref{equ2}) may be
rewritten as:
\begin{equation}\label{equ3}
\rho_{s}(0)~\Rightarrow~\rho_{s}(t)=\sum_{m}|C_{m}(t)|^{2}|m\rangle\langle{}m|.
\end{equation}
When the time $t \gg \tau_d$, the non-diagonal elements of the
reduced density matrix $\rho_s(t)$ vanish and the diagonal elements
achieve their equilibrium values. This effect of decoherence is
typical for all known quantum systems that induces an increase of
the system's entropy and the damping of quantum oscillations with
time \cite{Giulini,Leggett}.\\

A theoretical description of the evolution of the system from
$\psi(0)$ to $\psi(t)$ driven externally by the environment is
generally a very difficult problem. The case that the environment is
described by Boson fields has been extensively studied in the
context of Master equation approach, both with Markovian
\cite{Gardiner} or non-Markovian \cite{Shresta} approximations.
Although the master equation scheme can be used for a large number
of environments of possible types (phonon, photons, etc.)
\cite{Leggett}, however, the Master equation description is not
universally valid for all the models of environment and
fragile in some systems \cite{Frasca}.\\

Generally, if the Hamiltonian of the compound system is known, the
direct way to solve the decoherence problem is to follow the
evolution of the compound system over a substantial period of time.
By setting $\hbar=1$, the time dependent Schr\"odinger equation is:
\begin{equation}\label{equ4}
i\frac{\partial\psi(t)}{\partial~t}=\hat{H}\psi(t).
\end{equation}
Here $\hat{H}$ is the total Hamiltonian of the system plus the
environment. The equation (\ref{equ4}) can be decomposed into a set
of first-order ordinary differential equations with the initial
condition $\psi(0)$, and the total number of equations is the
dimension of the Hilbert space of the whole system, which is usually
very large. In principle, the set of equations can be solved by
convenient methods of ordinary differential equations such as
Predictor-Corrector method or Runge-Kutta method. However, direct
solution of the equations will cost too much computer resource due
to the large number of equations involved. Another scheme for
propagating equation (\ref{equ4}) is to expand the evolution
operator $U(t)=\exp(-i\hat{H}\Delta{}t)$ in a Taylor series, where
$\Delta{}t$ is the time step.
\begin{equation}
\exp(-i\hat{H}\Delta{}t)=1-i\hat{H}\Delta{}t+\cdots.
\end{equation}
It has been stated in Ref. \cite{TalEzer} that a numerical scheme
based on this expansion is not stable, because it does not conserve
the time reversal symmetry of Schr\"odinger equation. Variations of
the Taylor series have been proposed and used in calculations of
evolution of quantum systems \cite{Askar,Kosloff0}. Efficient and
stable simulation methods are needed to reduce the computation load
and to increase the simulation speed. \\

The polynomial expansion method has been used in the calculation of
dynamics and/or spectral properties of large quantum systems with
great success \cite{TalEzer,Weiffe,Kosloff,Silver}, Tal-Ezer and
Kosloff proposed the expansion in terms of the Chebyshev polynomials
and tested the method with the simple harmonic oscillator and the
problem of scattering from a surface, high accurate results were
obtained with an efficiency six times higher compared to the
conventional scheme \cite{TalEzer,Kosloff}. Silver and R\"oder used
the Chebyshev polynomial expansion in the calculation of density of
states of large sparse Hamiltonian matrix \cite{Silver}. A fast
evolution method based on the expansion of Chebyshev polynomial for
dynamical quantum systems was proposed and checked by Loh et al
\cite{Loh}. Dobrovitski et al extended the Chebyshev polynomial
expansion method in the study of a spin system interacting with a
spin-bath \cite{Dobrovitski1}, obtained the decoherence properties
of the system and showed the efficiency and accuracy of the method.
Since the Chebyshev polynomial is the most frequently used
orthogonal polynomial in most numerical approximation theories
\cite{cheney}, other kinds of orthogonal polynomials should also be
applicable in the evolution problems. The argument of Chebyshev
polynomial is bounded to the interval $[-1, +1]$, which is suitable
for systems with a bounded Hamiltonian, and for systems that only
bounded below, a cut off to the energy spectrum is inevitable in
order to use the method. However, it is well known that some of the
orthogonal polynomials, like Hermite polynomial and Laguerre
polynomial, do not limit their arguments to finite intervals.
Expansion in terms of these kinds of orthogonal polynomials may have
the merit in the unbounded systems. In this paper, we will explore
the efficiency and accuracy of methods based on all these orthogonal
polynomials. We constructed methods based on the Hermite and
Laguerre polynomials and found that the above two mentioned
orthogonal polynomials do have the required properties. The rest of
the paper is organized as follows. In Section II, we briefly review
the spin-bath model and the difficulty on getting its exact
solution; In Section III, three kinds of polynomial scheme will be
described for the expansion of the evolution operator; In Section
IV, we present the results of our numerical simulation; Finally, A
brief summary is given in the Section V.

\section{Hamiltonian} \label{hamiltonian}
Two systems are used in this study to test the numerical methods.
The first is a two spin-$1/2$ system coupled to a spin environment
and the second is a particle moving in a double well potential.\\

The spin Hamiltonian we used in testing our numerical schemes is the
one that used in reference
\cite{Dobrovitski1,Dobrovitski2,Prokof¡¯ev}. The system consists of
two spins-$1/2$ interacting antiferromagneticly, and the system
coupled to a bath of non-interacting spins-$1/2$. The Hamiltonian
can be written as:
\begin{equation}  \label{equ6}
H=2J \mathbf s_1 \cdot \mathbf s_2 + \sum_k A_k (\mathbf s_1 +
\mathbf s_2)\cdot \mathbf I_k.
\end{equation}
Here $\mathbf s_1$ and $\mathbf s_2$ are two spins with spin half
coupled by the coupling constant $J$, favoring the antiparallel
alignment, which constitute the system. The spins $\mathbf I_k$,
$k=1, 2, \cdots, N$ are $N$ spin half environment spins, interacting
with the system by Heisenberg coupling $A_k$, and do not interact
with each other. The coupling constant between the two system spins
is much larger than the couplings to the environment spins, $J\gg
A_k$. The couplings $A_k$ are uniformly distributed in an interval.
Both of the system spins and the environment spins can be
represented by Pauli matrices. \\

The Hilbert space of the whole system is $2^{N+2}$ dimensional when
the environment consists of $N$ spins. The basis state of the
environment can be chosen as the direct product of the single states
$\mid\uparrow\rangle$ or $\mid\downarrow\rangle$ for each spin
$\vec{I}_k$, here $\mid\uparrow\rangle$ and $\mid\downarrow\rangle$
are eigenstates of the square and $z$ components of each spin. For a
moderate size of the environment, say, $N=18$, we have to find an
exact solution to about $10^{6}$ differential equations. And when
$N$ is increased by one, the number of equations is doubled. For
this reason efficient algorithms are needed in the studies of the
evolution of this kind of problems, especially in the case of
decoherence where long time simulation is required to reach the
pointer state. The polynomial expansions based on both Chebyshev
polynomial \cite{Dobrovitski1} and Hermite polynomial \cite{Hu} are
very successful in this case.\\

The Hamiltonian for the double well potential is given by:
\begin{equation}\label{dW}
H=\frac{p^2}{2}-\frac{1}{2}\omega^2x^2+\lambda{}x^4.
\end{equation}
where we set $m=\hbar=1$. This model is very important in the
studies of critical phenomena and in the standard model of particle
physics when the variable $x$ is a scalar field. Here we take it to
be a simple yet non trivial model to test our numerical method.

\section{Polynomial Scheme}
\label{Poly}

The formal solution of the equation (\ref{equ4}) is:
\begin{equation}\label{equ5}
\psi(t)=e^{-i\hat{H}t}\psi(0)=U(t)\psi(0).
\end{equation}
The evolution operator $U(t)$ is an exponential functional of the
Hamiltonian operator $\hat{H}$ which is represented as a matrix in
the Hilbert space of $\psi$. The method of polynomial expansion is
to expand the evolution operator $U(t)$ in terms of the orthogonal
polynomials of Hamiltonian $\hat{H}$. The expansions in Chebyshev
polynomial and Hermite polynomial are presented in
\cite{Dobrovitski1} and \cite{Hu} respectively. We will briefly
introduce the Chebyshev and Hermite polynomial expansion and give
detailed derivation of expansion in terms of Laguerre polynomials
and check the efficiency of the method numerically.

\subsection{Chebyshev polynomial}
\label{Cheb}

The Chebyshev expansion of $U(t)$ given by Dobrovitski et al is:
\begin{equation}\label{equ7}
U(t)=\exp(-i\tau\tilde{H})=\sum^{\infty}_{k=0}c_{k}T_{k}(\tilde{H}),
\end{equation}
where $\tau=E_{0}t/2$ and $\tilde{H}=2\hat{H}/E_{0}$, $E_0$ is a
scale factor, $T_{k}$ are the Chebyshev polynomials:
$T_{k}(x)=\cos(k\arccos{}x)$. The reason that we change $\hat{H}$
into $\tilde{H}$ comes from the argument domain of $T_{k}(x)$, that
is $x\in[-1, 1]$. For our spins system, $\hat{H}$ is bounded above
and below, so that the scale factor $E_0$ can be determined in the
following way:
\begin{eqnarray*}
E_{\text {max}} & = & {\text {max}}\langle\psi|\hat{H}|\psi\rangle, \\
E_{\text {min}} & = & {\text {min}}\langle\psi|\hat{H}|\psi\rangle, \\
E_{0} & = & 2\,{\text {max}}(|E_{\text {max}}, E_{\text {min}}|).
\end{eqnarray*}

Using the orthogonal property of $T_{k}$, the expansion coefficients
$c_{k}$ of equation (\ref{equ7}) can be calculated as:
\begin{equation*}
c_{k}=\frac{a_{k}}{\pi}\int^{1}_{-1}\frac{T_{k}\exp(-ix\tau)}{\sqrt{1-x^{2}}}dx
=a_{k}(-i)^{k}J_{k}(\tau),
\end{equation*}
where $J_{k}(\tau)$ is the Bessel function of the $k$th order, and
$a_{0}=1$ when $k=0$ and $a_{k}=2$ when $k\geqslant1$. The series of
Chebyshev polynomials of Hamiltonian $\hat H$ can be calculated by the
recursion process:
\begin{eqnarray*}
T_{0}(\tilde{H}) & = & 1,\\
T_{1}(\tilde{H}) & = & \tilde{H},\\
T_{k+1}(\tilde{H}) & = &
2\tilde{H}T_{k}(\tilde{H})-T_{k-1}(\tilde{H}).
\end{eqnarray*}

\subsection{Hermite polynomial}
In order to obtain the expansion in terms of Hermite polynomials, we
start from its generating function \cite{Arfken}
\begin{equation}\label{equ8}
e^{-s^{2}+2sx}=\sum^{\infty}_{k=0}\frac{s^{k}}{k!}H_{k}(x).
\end{equation}
Where $H_{k}(x)$ denotes the Hermite polynomial of order $k$.
The evolution operator (\ref{equ5}) can be rearranged as
\begin{equation}\label{equ9}
e^{-i\hat{H}t}=e^{-(t/2\lambda)^{2}}e^{-(-it/2\lambda)^{2}+2\lambda\hat{H}(-it/2\lambda)}.
\end{equation}
The second part of the right hand side of equation (\ref{equ9}) is
identified to be the generating function of Hermit polynomial by
setting $x=\lambda\hat{H}$ and $s=-it/2\lambda$ in equation
(\ref{equ8}), where $\lambda$ is introduced for convenience. From
equation (\ref{equ8}) and (\ref{equ9}) we obtain the Hermite
expansion form of the exponential operator $U(t)$:
\begin{equation}\label{equ10}
e^{-i\hat{H}t}=e^{-(t/2\lambda)^{2}}
\sum^{\infty}_{k=0}\frac{(-i)^{k}}{k!}(t/2\lambda)^{k}H_{k}(\lambda\hat{H}).
\end{equation}
The formal solution $\psi(t)=\exp(-i\hat{H}t)\psi(0)$ then becomes:
\begin{eqnarray}\label{equ11}
\psi(t) &=& e^{-(t/2\lambda)^{2}}\sum^{\infty}_{k=0}\frac{(-i)^{k}}{k!}(t/2\lambda)^{k}\phi_{k}, \\
\phi_{k}&=& H_{k}(\lambda\hat{H})\psi(0). \nonumber
\end{eqnarray}
The Hermite polynomial of $H$ can be obtained by the following
recursive algorithm:
\begin{eqnarray*}
\phi_{0} & = & \psi_{0},\\
\phi_{1} & = & 2\lambda\hat{H}\psi_{0},\\
\phi_{k+1} & = & 2\lambda\hat{H}\phi_{k}-2k\phi_{k-1}.
\end{eqnarray*}
To discuss the convergence of the expansion, we consider the term when
$k$ is large. The Hermite polynomial may be replaced by its asymptotical
expression \cite{Arfken}:
\begin{equation}\label{equ:Herasy}
H_k(x)\approx2^{\frac{k+1}{2}}k^{\frac{k}{2}}e^{-\frac{k}{2}+\frac{x^2}{2}},
\cos\left(\sqrt{2k+1}x-\frac{k\pi}{2}\right).
\end{equation}
Substitute this into equation (\ref{equ10}) and using the Stirling's formula for the factorial,
\begin{equation}\label{stir}
 k! \approx \exp[k(\ln{}k-1)], \quad k \gg 1,
\end{equation}
the magnitude of the $k$th term in the expansion of equation (\ref{equ10}) for large $k$ is:
\begin{equation}\label{equ:expan}
\frac{(t/2\lambda)^{k}}{k!}H_k(\lambda\hat{H}) \approx
\frac{(t/\lambda)^{k}}{2^ke^{k(\ln{}k-1)}}2^{\frac{k+1}{2}}k^{\frac{k}{2}}
e^{-\frac{k}{2}+\frac{\lambda^2\hat{H}^2}{2}}\cos\left(\sqrt{2k+1}\lambda\hat{H}-\frac{k\pi}{2}\right).
\end{equation}
The physically meaningful Hamiltonian should always be bounded
below, and for every evolution problem, the spectrum of the system
has a maximum value determined by the initial state, which is in the
order of the total energy of the initial state. If we set a maximum
energy $E_m$, a few times of the total energy, then the states with
energy larger than this maximum will not enter the calculation, and
we have a natural energy cut off of the problem, the $E_m$. Then we
can replace $\hat{H}$ in equation (\ref{equ:expan}) with $E_m$ to
estimate the condition of the convergence of the expansion.
\begin{eqnarray*}
\frac{(t/2\lambda)^{k}}{k!}H_k(\lambda{}E_m) & \leq &
2^{-\frac{k-1}{2}}\exp\left[-\frac{k}{2}\ln{}k+\frac{k}{2}+\frac{\lambda^2E_m^2}{2}
 +k\ln\left(\frac{t}{\lambda}\right)\right] \\
 & = & 2^{-\frac{k-1}{2}}\exp\left\{-\frac{k}{2}\left[\ln{}k-\ln{}e+
 \ln\left(\frac{t}{\lambda}\right)^{-2}-\frac{\lambda^2E_m^2}{k}\right]\right\} \\
 & =&
 2^{-\frac{k-1}{2}}\exp\left\{-\frac{k}{2}\left[\ln\left(\frac{k\lambda^2}{et^2}\right)-
 \frac{\lambda^2E_m^2}{k}\right]\right\}.
\end{eqnarray*}
From this expression we see that if
\begin{equation*}
\ln\left(\frac{k\lambda^2}{et^2}\right)-\frac{\lambda^2E_m^2}{k}\geq0,
\end{equation*}
or the time step $t$ satisfies
\begin{equation}\label{dtofH}
t\leq\sqrt{\frac{k}{e}}\lambda\exp\left(-\frac{\lambda^2E_m^2}{2k}\right).
\end{equation}
The $k$th term is not larger than $2^{-\frac{k-1}{2}}$, then the
summation is convergent. In the numerical calculation given below,
we set $\lambda=1/2$.

\subsection{Laguerre polynomial}
\label{Lag}

The expansion in terms of Laguerre polynomials can also be derived
from its generating function \cite{Arfken}:
\begin{equation}\label{equ12}
(1-s)^{-\alpha-1}e^{xs/(s-1)}=\sum^{\infty}_{k=0}L^{\alpha}_{k}(x)s^{k},\quad(|s|<1).
\end{equation}
where $\alpha$ distinguishes different types of Laguerre
polynomials. By setting $s=it/(\lambda+it)$ and $x=\lambda\hat{H}$,
we get the Laguerre polynomial expansion as:
\begin{eqnarray}\label{equLa}
\psi(t) &=& \left(\frac{\lambda}{\lambda+it}\right)^{\alpha+1},
\sum^{\infty}_{k=0}\left(\frac{it}{\lambda+it}\right)^k\phi_{k}, \\
\phi_{k}&=& L^{\alpha}_k(\lambda\hat{H})\psi(0). \nonumber
\end{eqnarray}
The recursion relation of Laguerre polynomials are
\begin{eqnarray}
L^{\alpha}_{0}(x) & = & 1\\ \nonumber
L^{\alpha}_{1}(x) & =
&\alpha+1-x\\ \nonumber
(k+1)L^{\alpha}_{k+1}(x) & = &
(2k+\alpha+1-x)L^{\alpha}_{k}(x)-(k+\alpha)L^{\alpha}_{k-1}(x). \label{equ14}
\end{eqnarray}
From the relation we obtain the Laguerre polynomial expansion of
Hamiltonian $\hat{H}$ as:
\begin{eqnarray}
\phi^{\alpha}_{0} & = & \psi(0)\\ \nonumber \phi^{\alpha}_{1} & =
&(\alpha+1-\lambda\hat H) \psi(0)\\ \nonumber
(k+1)\phi^{\alpha}_{k+1} & = & (2k+\alpha+1-\lambda\hat
H)\phi^{\alpha}_{k}-(k+\alpha)\phi^{\alpha}_{k-1}. \label{equ15}
\end{eqnarray}
Different $\alpha$ gives different choice of the algorithms, the
domain of $\alpha$ is in the interval of $(-1,\infty)$. In the
calculation of the spin bath Hamiltonian we use $\alpha=-1/2$ and
set the parameter $\lambda=1$ for convenience. For other kinds of
Hamiltonian different values of $\alpha$ may be used to attain
higher efficiency and accuracy.\\

The convergency of the expansion of equation (\ref{equLa}) is
guaranteed by the relationship between Laguerre polynomial and
Hermite polynomial \cite{Arfken}:
\begin{equation}\label{equ:LH}
L_k^{-\frac{1}{2}}(x)=\frac{(-1)^k}{2^{2k}k!}H_{2k}(\sqrt{x}).
\end{equation}
Substituting equation (\ref{equ:Herasy}), (\ref{stir}) and
(\ref{equ:LH}) into the expansion term
$\left(\frac{it}{1+it}\right)^kL^{-\frac{1}{2}}_k(\hat{H})$ and
replacing $\hat{H}$ with $E_m$, the total energy of the initial
state, we could estimate its asymptotical absolute value by such a
procedure:
\begin{eqnarray*}
\left|\left(\frac{it}{1+it}\right)^kL^{-\frac{1}{2}}_k(E_m)\right| &
\approx &
\left(\frac{t^2}{\lambda^2+t^2}\right)^{k/2}\frac{1}{2^{2k}e^{k(\ln{}k-1)}}
2^{\frac{2k+1}{2}}2^kk^ke^{-k+\frac{E_m}{2}}\cos\left(\sqrt{2k+1}E_m-\frac{k\pi}{2}\right) \\
& \leq &
2^{\frac{1}{2}}\left(\frac{t^2}{\lambda^2+t^2}\right)^{k/2}e^{\frac{E_m}{2}}
\\ & = & \exp\left\{-k/2\left[\ln\left(\frac{1+t^2}{t^2}\right)-\frac{E_m+\ln2}{k}\right]\right\}
\end{eqnarray*}
For large $k$, and the suitably chosen time step
\begin{equation}\label{dtofH}
t<\left[\exp\left(\frac{E_m+\ln2}{k}\right)-1\right]^{-\frac{1}{2}},
\end{equation}
the terms approach to zero exponentially.\\

It should be noted that the energy cut off $E_m$ is only used here
for convergence proof. In practical calculations, we do not need to
specify this cut off and the time step is chosen by test and error.
\\

Comparing to the Chebyshev expansion, the methods of Hermite
polynomial and Laguerre polynomial have an obvious advantage that no
scaling to the Hamiltonian is needed. So that these expansions may
have wider applications. On the other hand, the recurrence relation
for both Hermite polynomial and Laguerre polynomial is not
numerically absolute stable as compared to the recurrence relation
of Chebyshev polynomial, which is marginal stable \cite{Press}. This
fact limits the number of terms in the expansion to some value
$k_{\text{max}}$, the effect of numerical instability has little
effect for $k< k_{\text{max}}$ and the effect starts to show up
beyond this cut off. In the practical calculations the
$k_{\text{max}}$ may be chosen to be $30$, and the time step is set
up accordingly with a specified error tolerance to get convergent
results. The calculation schemes presented here are very general and
are not dependent on the specific form of the Hamiltonian, however,
the applicability should be tested for each kind of Hamiltonian
before it can be used in practical simulations. The efficiencies of
the three kinds of polynomial expansion are almost the same from our
numerical calculation, careful comparison reveals that for the
current models the Laguerre expansion with $\alpha=1/2$ is a little
bit faster than the others.

\section{Numerical simulation}

\subsection{Test of the spin model}
The efficiency of the Chebyshev expansion over conventional method
of calculation has already given by
\cite{Dobrovitski1,Dobrovitski2}. In this section we checked
numerically the efficiency of three kinds of polynomial expansions
by comparing the performance among the three expansions as well as
with the {\it predictor-corrector (P-C)} and {\it Runge-Kutta (R-K)}
methods to the spin bath Hamiltonian given in section
\ref{hamiltonian}. We calculated two particular variables using the
Hamiltonian: (i) the $z$-component oscillation of any one of the
center spins, i.e. $s^{z}_{i}$, $i=1$ or $2$, which demonstrated the
decoherence rate of the system; (ii) the time dependence of von
Neumann's entropy, i.e. $S_{vN}=-Tr\rho\ln\rho$, which characterizes
the entanglement degree of the state of the system \cite{Zurek1981}.
We use the same parameters as used in
\cite{Dobrovitski1,Dobrovitski2}, the exchange strength $J=16.0$,
$A_{k}$ are uniformly distributed between $0$ and $0.5$; The initial
condition of the system is
$|\psi(0)\rangle=\mid\uparrow\downarrow\rangle$ or written as
$|10\rangle$, and the environment is a normalized linear
superposition of the product states of $N$ spins with random
coefficients. The time step is chosen as $\Delta{}t=0.036$, which is
determined by the compromise of convergence requirement,
$\Tr(\rho)=1$, and the speed of computation. All of the three
schemes are implemented and tested, the results are consistent with
those given by \cite{Dobrovitski1,Dobrovitski2}. We also did the
calculation with the two widely used ordinary differential equation
solver, the {\it predictor-corrector (P-C)} and {\it Runge-Kutta
(R-K)} methods. At the request of stability and speed, the time step
in these two methods is almost $1/10$ of that in polynomial scheme.
We found that the calculation cost of the three polynomial expansion
schemes are very close to each other, with the Laguerre polynomial
expansion is slightly faster, and the results are practically the
same. So we only give the datum thereinafter obtained by Laguerre
polynomial expansion in the following.

\begin{figure}[htbp]
\centering
\subfigure[Oscillation of $s^{z}_{1}(t)$ ]{
\label{standard:sz}
\includegraphics[width=3in]{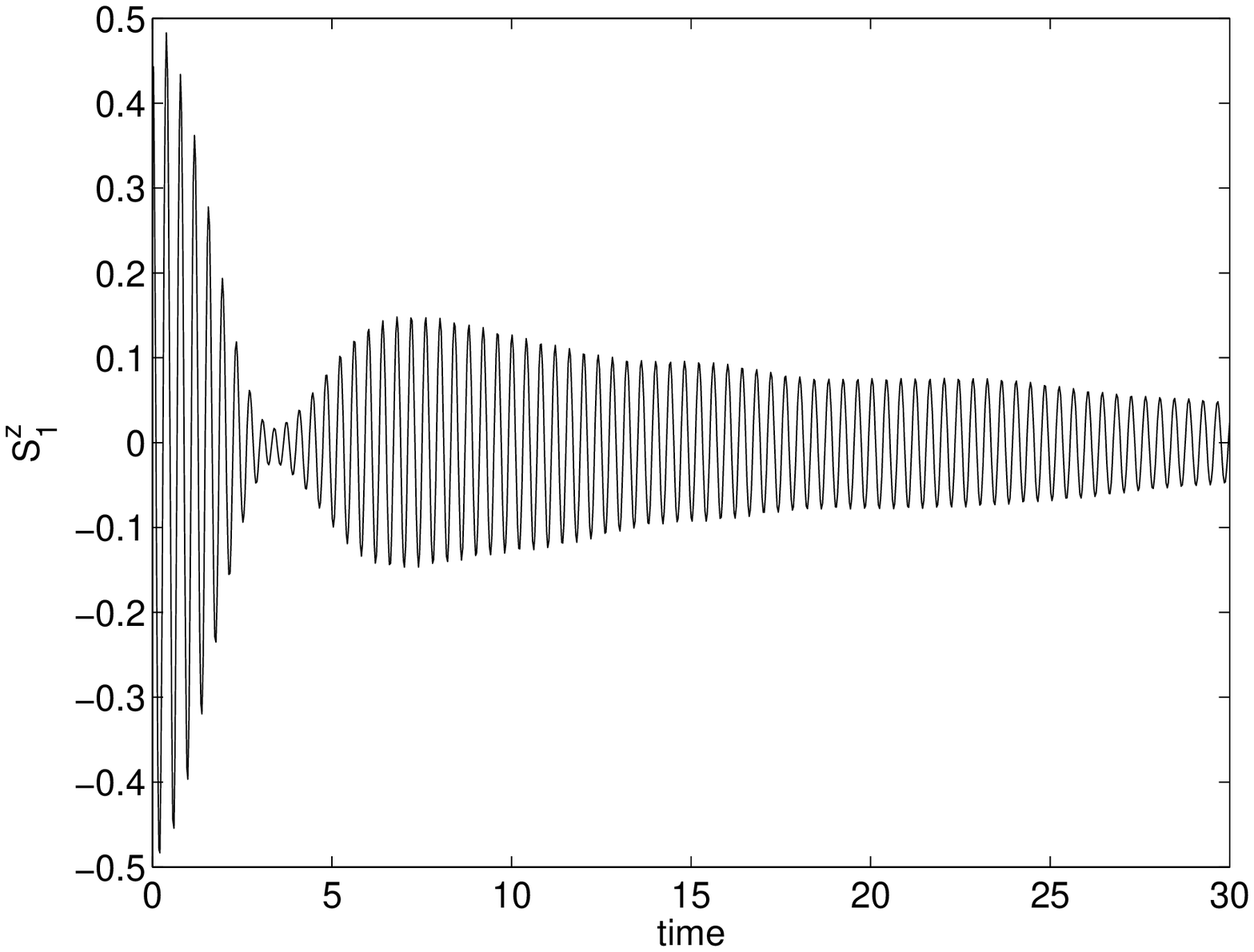}}
\subfigure[Evolution of entropy $S_{vN}(t)$]{
\label{standard:entropy}
\includegraphics[width=3in]{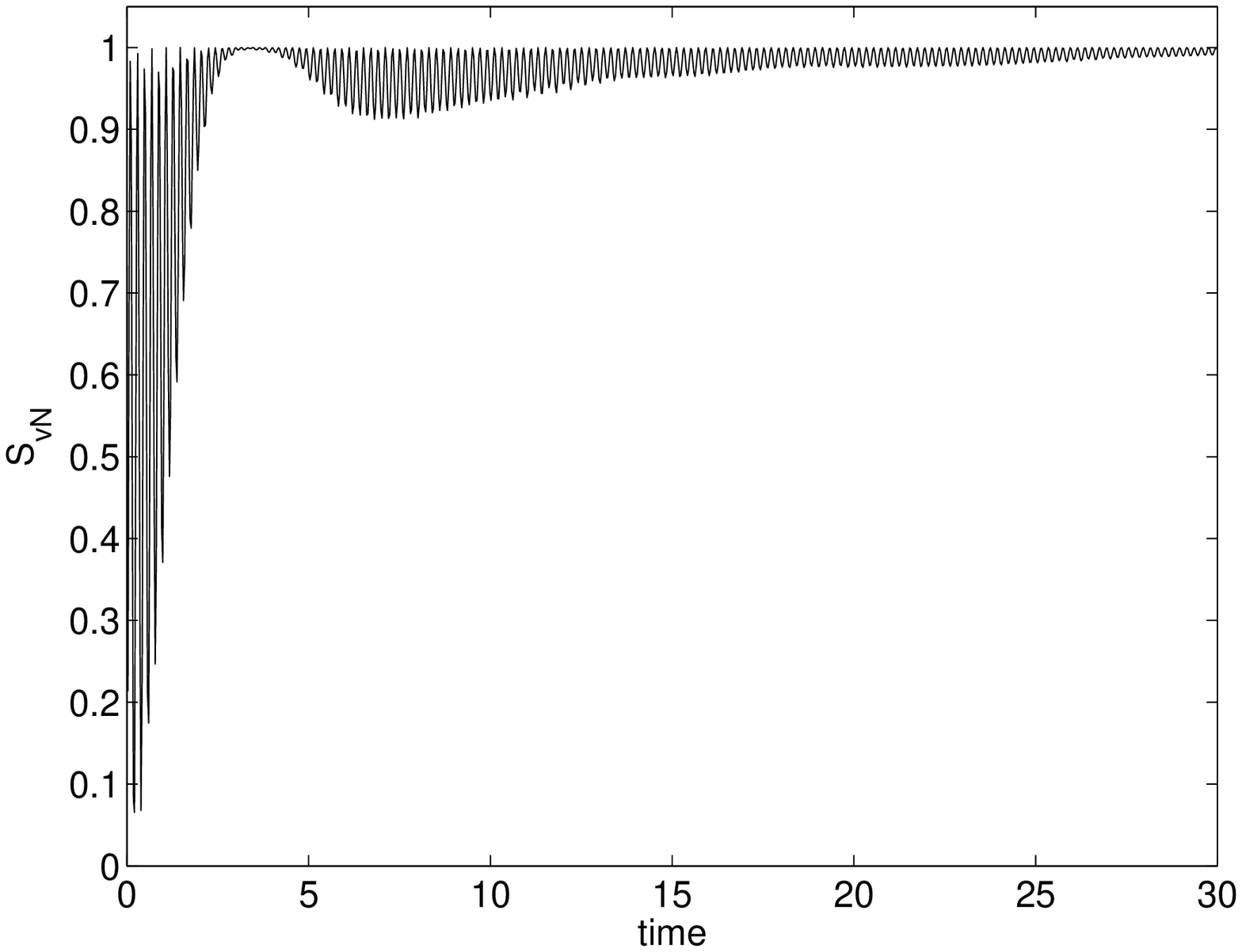}}
\caption{Decoherence of two coupled spins by a spin bath
calculated by Laguerre method, the parameters are: $J=16$, $N=12$,
the tolerance in obtained this figure is set to be $10^{-6}$.}
\label{standard}
\end{figure}

\begin{table}
\caption{Comparison of the \textbf{R-K} method with the polynomial
scheme (abbreviated as \textbf{P-S})for the problem of decoherence
of spin-bath}
\begin{center}
\begin{tabular}{c|c|c|c|c|c}
\hline\hline {\sl Scheme}&{\sl $\Delta{}t$}&{\sl No. of bath-spin
}&{precision}&{$t$}&{CPU time}\\ \hline
R-K & $0.0036$ & $4$ & $10^{-6}$ & $9000\Delta{}t$ & $2$ sec \\
P-S & $0.036$  & $4$ & $10^{-6}$ & $900\Delta{}t$  & $2$ sec \\
\hline
R-K & $0.0036$ & $8$ & $10^{-6}$ & $9000\Delta{}t$ & $406$ sec \\
P-S & $0.036$  & $8$ & $10^{-6}$ & $900\Delta{}t$  & $ 50$ sec \\
\hline
R-K & $0.0036$ & $10$ & $10^{-6}$ & $9000\Delta{}t$ & $2065$ sec \\
P-S & $0.036$  & $10$ & $10^{-6}$ & $900\Delta{}t$ & $242$ sec\\
\hline\hline
\end{tabular}
\end{center}
\label{comp}
\end{table}

Figure \ref{standard} are results of the oscillation of
$s^{z}_{1}(t)$ and von Neumann's entropy $S_{vN}(t)$ of the
spin-bath Hamiltonian with parameters given in the figure caption.
The results are obtained by the Laguerre polynomial expansion method
and are consistent with results by other methods we tested and those
reported in literature \cite{Dobrovitski1,Dobrovitski2}.\\

The comparison between computation costs of different methods with
the same error tolerance listed in Table \ref{comp}. From the table
we see that: (i) When $N$ is very small, it is hard to distinguish
the calculation speed of the two kinds of numerical computation
methods; (ii) In general, the speed of polynomial scheme is about
$8$ times as fast as that of the direct solution methods,
i.e, the Runge-Kutta (R-K) methods (the corresponding datum of
predictor-corrector method are almost the same as R-K); (iii)
With increasing $N$, the speed advantage becomes more evident. All
the data reported here are obtained by a micro computer with Intel
Pentium M Banias Processor 1400MHz, Memory 256M.

\subsection{The double well model with Laguerre polynomial scheme}

\begin{figure}[htbp]
  \centering
\includegraphics[scale=0.7]{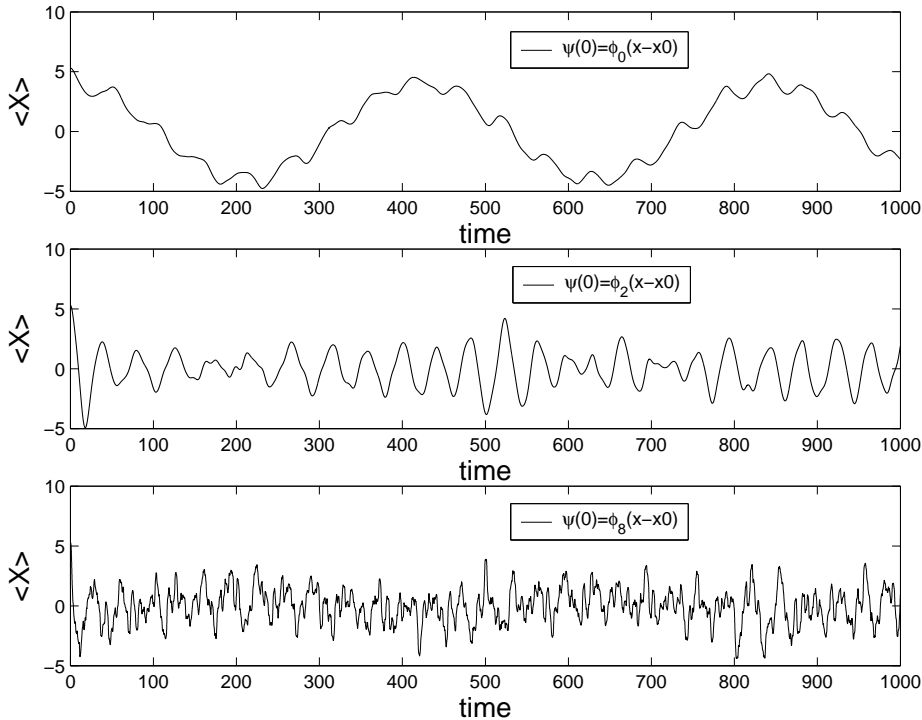}\\
\caption{The time evolution of $\langle{}x\rangle$ of three cases:
(a), $\psi(0)=\phi_0(x-x_0)$; (b), $\psi(0)=\phi_2(x-x_0)$; (c),
$\psi(0)=\phi_8(x-x_0)$. All of them are calculated in the condition
of $\lambda/\omega=0.0013$.}\label{fig:X028}
\end{figure}

\begin{figure}[htbp]
  \centering
\includegraphics[scale=0.7]{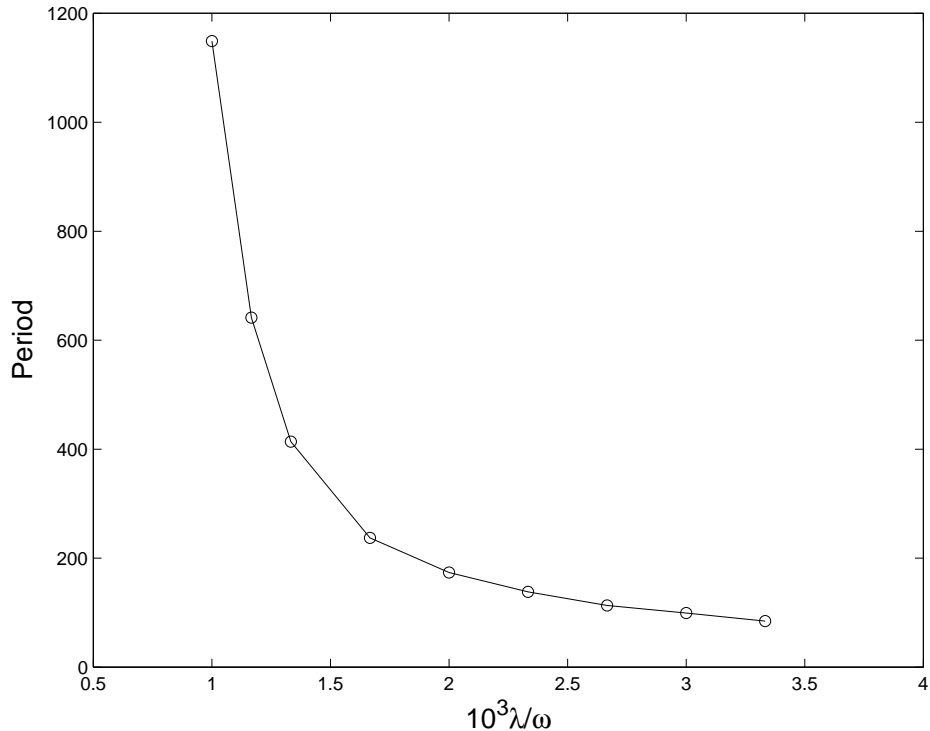}\\
\caption{The period of different $\lambda/\omega$ with the same
initial state $\psi(0)=\phi_0(x-x0)$}\label{fig:period}
\end{figure}

\begin{figure}[htbp]
  \centering
\includegraphics[scale=0.7]{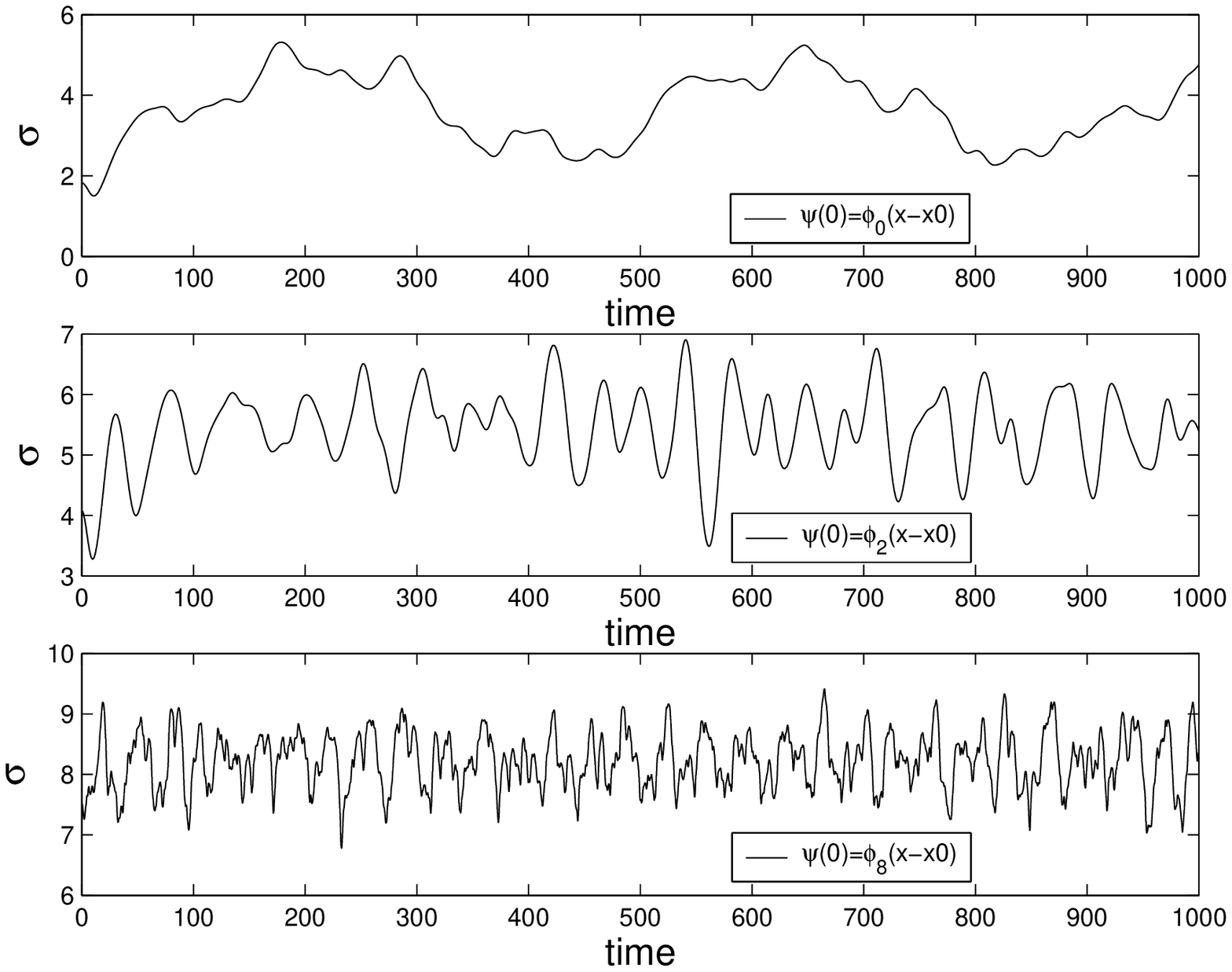}\\
\caption{The time evolution of Standard deviation of coordinate
$\sigma=(\langle{}x^2\rangle-\langle{}x\rangle^2)^{1/2}$ of three
cases: (a), $\psi(0)=\phi_0(x-x_0)$; (b), $\psi(0)=\phi_2(x-x_0)$;
(c), $\psi(0)=\phi_8(x-x_0)$. All of them are calculated in the
condition of $\lambda/\omega=0.0013$.}\label{fig:VAR}
\end{figure}

The Laguerre polynomial expansion scheme can easily be extended into
the studies of continuous quantum systems. As an illustration, we
used it in the calculation of the time-evolution of a given wave
function packet in the double well system. The initial state
prepared as one of the eigenstates of a harmonic oscillator with
unit mass and frequency $\omega$, centered at the bottom of the
right well, $x_0=\omega/\sqrt{4\lambda}$. That is:
\begin{equation}  \label{initial}
\psi(0)=\left(\frac{\sqrt{\omega}}{\sqrt{\pi}2^mm!}\right)^{1/2}
H_m(\sqrt{\omega}(x-x_0))\exp[-\omega(x-x_0)^2/2].
\end{equation}
$H_m(x)$ is the Hermite polynomial of the $m$th order.\\

In order to use the Laguerre polynomial expansion scheme in the
evaluation of the time evolution, we expand the state of the system
by a complete basis state. In principle, any complete basis can be
used in this calculation, however, a better choice of the basis will
greatly reduce the computation efforts and obtain high accuracy
results. In this study we use the eigenstates of a simple harmonic
oscillator, $\phi_n(x)$, $n=0,1,\cdots\infty$, abbreviated as
$|n\rangle$ as as the expansion basis. The Hamiltonian of the simple
harmonic oscillator that defines the basis is
\begin{equation}\label{sh}
h = \frac12 p^2+\frac12 \omega^2x^2.
\end{equation}
This is not necessarily the optimized basis, however, calculation
shows that it is pretty good in this problem. \\

By introduction of the creation operator $a^+$ and annihilation
operator $a$, the matrix elements of the double-well Hamiltonian can
easily be evaluated. The coordinate $x$ and momentum $p$ can be
represented in terms of the operator $a^+$ and $a$:
\begin{eqnarray}\label{xp2a}
x &=& \sqrt{\frac{1}{2\omega}}(a^+ +a),\\ \nonumber  p &=&
i\sqrt{\frac{\omega}{2}}(a^+-a).
\end{eqnarray}
The action of $a^+$ and $a$ on $|n\rangle$ are:
\begin{eqnarray}\label{cran1}
a|n\rangle &=& \sqrt{n}|n-1\rangle,\\ \label{cran2}  a^+|n\rangle
&=&\sqrt{n+1}|n+1\rangle, \\ \nonumber h|n\rangle &=&
\omega\left(n+\frac{1}{2}\right)|n\rangle.
\end{eqnarray}
In the $a^+$ and $a$ representation, the double-well Hamiltonian
(\ref{dW}) becomes:
\begin{equation} \label{dw2}
H=-\frac{1}{2}\omega[(a^+)^2+a^2]+\frac{\lambda}{4\omega^2}(a^++a)^4.
\end{equation}
By using (\ref{cran1}), the matrix elements of (\ref{dw2}) can
easily be obtained. And the matrix form of the Hamitonian can be
substituted directly in the Laguerre polynomial expansion scheme
providing a suitable cut off of the states is specified. In our
calculation, we cut off the states at $n=49$, at which in all cases
we studied are already convergent. The initial state $\psi(0)$ in
the calculation is also expanded in terms of the $|n\rangle$.
When $m=0$ in (\ref{initial}), the expansion is:
\begin{eqnarray*}
\psi(0) &=&
\exp\left(-\frac{1}{2}\alpha^2_0\right)\sum_{n=0}^N\frac{\alpha^n_0}{\sqrt{n!}}|n\rangle, \\
 \alpha_0 &=& x_0\sqrt{\frac{\omega}{2}}.
\end{eqnarray*}
For other value of $m$ in (\ref{initial}), the coefficients of the
expansion can easily be evaluated numerically.\\

Using the Laguerre polynomial scheme, we calculated the average
position $\langle{}x\rangle$ and the variation
$\sigma=\left(\langle{}x^2\rangle-\langle{}x\rangle^2\right)^{1/2}$.
Figure \ref{fig:X028} plotted the evolution of the average position
$\langle{}x\rangle$ with time. The initial states are the
eigenstates of simple harmonic centered at the right well of the
double well potential. For the state of $\phi_0(x-x_0)$, which is
located at the $x_0$ initially, it oscillates back and forth with
time. From figure \ref{fig:X028}(a) we see clearly the periodic
motion and the period can easily be identified. The period depends
on the value of $\lambda/\omega$. Smaller $\lambda/\omega$
corresponding to a deeper well and thus a longer period. Figure
\ref{fig:period} plots the period as a function of the ratio
$\lambda/\omega$, which is decreasing monotonically as expected. For
the states of higher energies, though the initial state is also
localized at the right potential well, the average position no
longer follows a periodic oscillation between the two wells,
instead, the particle spends most of the time moving around the
center of the potential. Figure \ref{fig:VAR} are plots of the
variation of the position, $\sigma=
\left(\langle{}x^2\rangle-\langle{}x\rangle^2 \right)^{1/2}$, as a
function of time, which represents the width of the corresponding
wave pocket. From the figure, we see that for the low energy state
$\phi_0(x-x_0)$, the width is typically $4$ as can be seen in the
figure, smaller than the total width of the potential at the average
energy of $\phi_0(x-x_0)$, which is about $10$, and it looks like a
wave pocket bouncing about. The energy of the state $\phi_0(x-x_0)$
for the parameters chosen is $-0.0390$, slightly lower than the
height of the middle peak of the potential. The movement of the
center of the particle between the two wells is a case of quantum
tunneling. In the higher energy cases, the wave pocket spends most
of the time oscillating around the center of the potential well and
there is no well defined period can be found. \\

\begin{figure}[htbp]
  \centering
\includegraphics[scale=0.7]{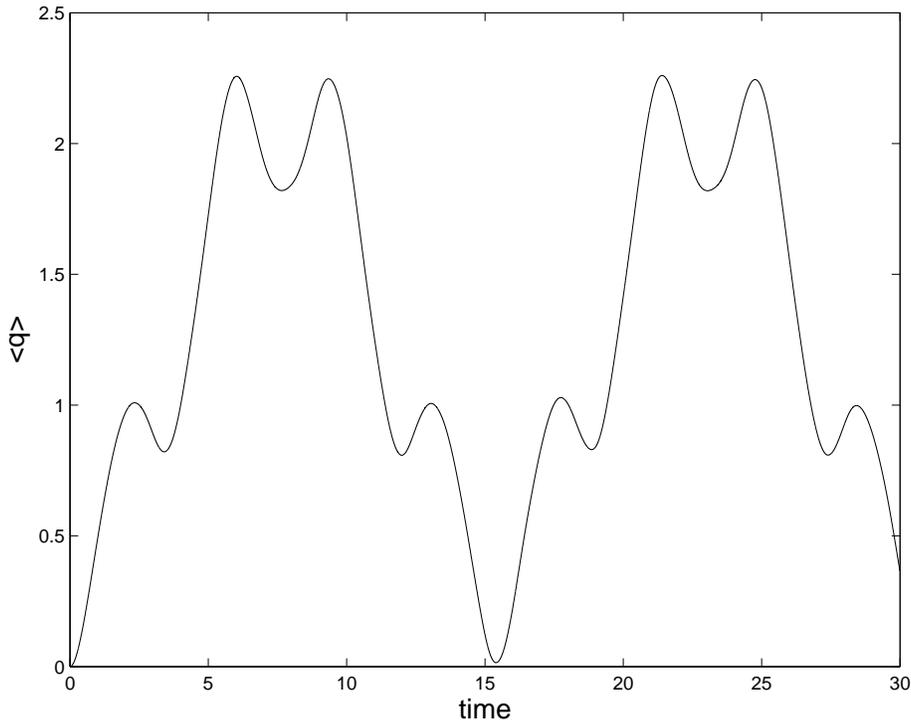}\\
\caption{Time dependence of $\langle{}q\rangle$ with
$\beta=2.5$}\label{fig:ref}
\end{figure}

A similar problem was studied by Bender et al many years ago
\cite{Carl}. If we transform the $x$ coordinate to $q$ according to
$q=x+\beta/2$ and set $\omega=\sqrt{8.0}$, the equation (\ref{dW})
is changed into
\begin{equation}
H=\frac{1}{2}p^2+4q^2(q-\beta)^2/\beta^2
\end{equation}
which is exactly the equation (1) in reference \cite{Carl}. We use
the same initial conditions as used in \cite{Carl} to calculate
$\langle{}q\rangle$ by our scheme (Here the number of energy
eigenstates $N$ is truncated to $32$, which is sufficient for
convergent). The result is given in figure \ref{fig:ref}, which is
the same as figure 1 in \cite{Carl}. The calculation time for this
figure is only about $4$ seconds in a personal computer of
Pentium(R) 4 CPU 2.60GHz, Memory 512M.

\section{Summary}
In summary, we proposed a new polynomial scheme, the Laguerre
polynomial expansion scheme, and tested its validity and efficiency
by means of the spin bath model and a continuous double-well model.
The obvious merit of this scheme compared to the Chebyshev
polynomial expansion scheme is that no scaling to Hamiltonian is
required, which means that a priori knowledge of the lower and upper
bounds of the Hamiltonian is not needed. On the other hand, the
computation efficiency and accuracy of the method are basically the
same as the Chebyshev polynomial expansion scheme. \\

We have also made use of the Laguerre expansion scheme in other
kinds of model systems to study the affection of the intra-bath
entanglement on the decoherence of the center spins. The method is
also as efficient and accurate in those models as it was in the
current spin bath model. The results will be reported in separate
publications.\\

This work is supported by the National Nature Science Foundation of China under grant
\#10334020 and \#90103035 and in part by the National Minister of Education
Program for Changjiang Scholars and Innovative Research Team in University.

\end{document}